# Renormalization of Massive Lattice Fermions

A.S. Kronfeld[a] and B.P. Mertens[b]

[a]Theoretical Physics Group, Fermi National Accelerator Laboratory, Batavia, Illinois, USA

[b]Enrico Fermi Institute and Department of Physics, University of Chicago, Chicago, Illinois, USA

The renormalization of a general action for massive lattice fermions is discussed. The analysis applies for all $m_q a$. Preliminary results for the self energy at one loop in perturbation theory are presented.

## 1. INTRODUCTION

This paper is a progress report of efforts to calculate the renormalization of fermion masses and bilinear currents in one-loop perturbation theory [1]. When these calculations are finished, they will permit a determination of heavy-quark masses, they will give one-loop predictions for the tuning of improvement parameters in the action, and they will give a one-loop guide for extrapolating matrix elements to the continuum.

After specifying a general action in sect. 2, sect. 3 sketches an all-orders derivation for mass and wavefunction renormalization in terms of the fermion self energy. We have the self energy to one-loop for the simplest action, and present results with and without tadpole improvement. Sect. 4 discusses current renormalization.

## 2. THE ACTION

Consider the action $S = S_0 + S_B + S_E + \cdots$,

$$\begin{aligned} S_0 = m_0 &\int \bar{\psi}(x)\psi(x) - \tfrac{1}{2}ar\zeta\int\bar{\psi}(x)\triangle^{(3)}\psi(x) \\ &+\zeta\int\bar{\psi}(x)\boldsymbol{\gamma}\cdot\boldsymbol{D}\psi(x) \\ &+\tfrac{1}{2}\int\bar{\psi}(x)[(1+\gamma_0)D_0^- - (1-\gamma_0)D_0^+]\psi(x), \end{aligned} \tag{1}$$

$$S_B = -\tfrac{i}{2}ac_B\zeta\int\bar{\psi}(x)\boldsymbol{\Sigma}\cdot\boldsymbol{B}(x)\psi(x), \tag{2}$$

$$S_E = -\tfrac{1}{2}ac_B\zeta\int\bar{\psi}(x)\boldsymbol{\alpha}\cdot\boldsymbol{E}(x)\psi(x). \tag{3}$$

Special cases are the Wilson action [2], which sets $r = \zeta = 1$, $c_B = c_E = 0$, and the Sheikholeslami-Wohlert action [3], which sets $r = \zeta = 1$, $c_B = c_E$. To remove lattice artifacts in general the parameters $m_0 a$, $r$, $\zeta$, $c_B$ and $c_E$ must all be adjusted [4]. In a non-relativistic setting, however, it is enough to adjust $m_0 a$, $c_B$, and $c_E$ [4].

## 3. THE SELF ENERGY

The self energy $\Sigma(p)$ is related to the momentum-space propagator by

$$S^{-1}(p) = S_0^{-1}(p) - \Sigma(p), \tag{4}$$

where $S_0(p)$ is the free propagator. In perturbation theory $\Sigma(p)$ is the sum of all one-particle irreducible graphs. The $p_0$-Fourier transform $C(t,\boldsymbol{p}) = (2\pi)^{-1}\int dp_0\, e^{ip_0 t}S(p)$ obeys

$$C(t,\boldsymbol{p}) = Z_2(\boldsymbol{p})e^{-E_{\boldsymbol{p}}|t|}Q + \cdots, \tag{5}$$

where $E_{\boldsymbol{p}}$ is the energy of a one fermion state with momentum $\boldsymbol{p}$, and $Q$ is a Dirac matrix satisfying $(Q\gamma_0)^2 = Q\gamma_0$. The $\cdots$ denote multi-particle states, which are irrelevant here.

The self energy has the decomposition

$$\Sigma(p) = i\sum_\mu \gamma_\mu \sin p_\mu a\, A_\mu(p) + C(p) \tag{6}$$

in Dirac matrices. For a Euclidean invariant cutoff $C$ and $A_\mu = A$ $\forall\mu$ are functions of $p^2$ only. With the lattice cutoff, however, they are constrained only by (hyper)cubic symmetry. For emphasis it is convenient to write, say, $C(p_0,\boldsymbol{p})$.

To obtain an expressions for $E_{\boldsymbol{p}}$ and $Z_2$ one carries out the $p_0$ integration with the residue theorem. For arbitrary $\boldsymbol{p}$ the energy $E_{\boldsymbol{p}}$ is the solution of the implicit equation

$$1 + m_0 a + \tfrac{1}{2} r\zeta \boldsymbol{p}^2 a^2 - C = \cosh Ea + (1 - A_0)\sqrt{1 - \mathcal{P}^2}\,\sinh Ea, \tag{7}$$

for $E$. The abbreviation

$$\mathcal{P}^2 = \frac{\sum_j (\zeta - A_j)^2 \sin^2 p_j a}{(1 - A_0)^2 \sinh^2 Ea}. \tag{8}$$

Here the self-energy functions $A_\mu(p_0, \boldsymbol{p})$ and $C(p_0, \boldsymbol{p})$ are evaluated at $p_0 = iE$. The solution of eq. (7), $E = E_{\boldsymbol{p}}$, defines the (lattice-distorted) mass shell of the fermion. The residue is

$$\begin{aligned} Z_2^{-1}(\boldsymbol{p}) = {} & (1 - A_0)\cosh E_{\boldsymbol{p}} a - \dot{C}\sqrt{1 - \mathcal{P}^2} \\ & + (\sqrt{1 - \mathcal{P}^2} + \dot{A}_0)\sinh E_{\boldsymbol{p}} a \\ & + \frac{\sum_j \dot{A}_j (\zeta - A_j)\sin^2 p_j a}{(1 - A_0)\sinh E_{\boldsymbol{p}} a}. \end{aligned} \tag{9}$$

The notation $\dot{f} = (ia)^{-1}(df/dp_0)$. In eq. (9) the self-energy functions $A_\mu(p_0, \boldsymbol{p})$ and $C(p_0, \boldsymbol{p})$ are evaluated on shell, i.e. $p_0 = iE_{\boldsymbol{p}}$.

The $\boldsymbol{p}$ dependence of $Z_2(\boldsymbol{p})$ is an artifact of the lattice cutoff. An acceptable definition of the wavefunction renormalization constant is

$$Z_2^{-1} = e^{M_1 a} - A_0 \cosh M_1 a + \dot{A}_0 \sinh M_1 a - \dot{C} \tag{10}$$

at $\boldsymbol{p} = \boldsymbol{0}$, where $M_1 \equiv E_{\boldsymbol{0}}$ is the (all-orders) rest mass of the fermion.

Eq. (7) at $\boldsymbol{p} = \boldsymbol{0}$ determines the rest mass via

$$e^{M_1 a} = 1 + m_0 a + A_0 \sinh M_1 a - C \tag{11}$$

and the dynamic mass $M_2 = (d^2 E_{\boldsymbol{p}}/dp_1^2)^{-1}_{\boldsymbol{p}=\boldsymbol{0}}$ via

$$\begin{aligned} \frac{e^{M_1 a}}{M_2 a} = {} & r\zeta + \frac{(\zeta - A_1)^2}{(1 - A_0)\sinh M_1 a} \\ & + \frac{d^2}{dp_1^2}\Big[A_0(iE_{\boldsymbol{p}}, \boldsymbol{p})\sinh E_{\boldsymbol{p}} a - C(iE_{\boldsymbol{p}}, \boldsymbol{p})\Big]. \end{aligned} \tag{12}$$

A total $p_1$-derivative includes an explicit part and an implicit part through the $E_{\boldsymbol{p}}$ dependence. In eqs. (11) and (12) the self-energy functions and derivatives are evaluated at $p_0 = iM_1$ and $\boldsymbol{p} = \boldsymbol{0}$.

For a massless fermion, $M_1 = M_2 = 0$. The bare mass that induces $M_1 = 0$ obeys

$$m_{0c} a = C(0, \boldsymbol{0}; m_{0c} a). \tag{13}$$

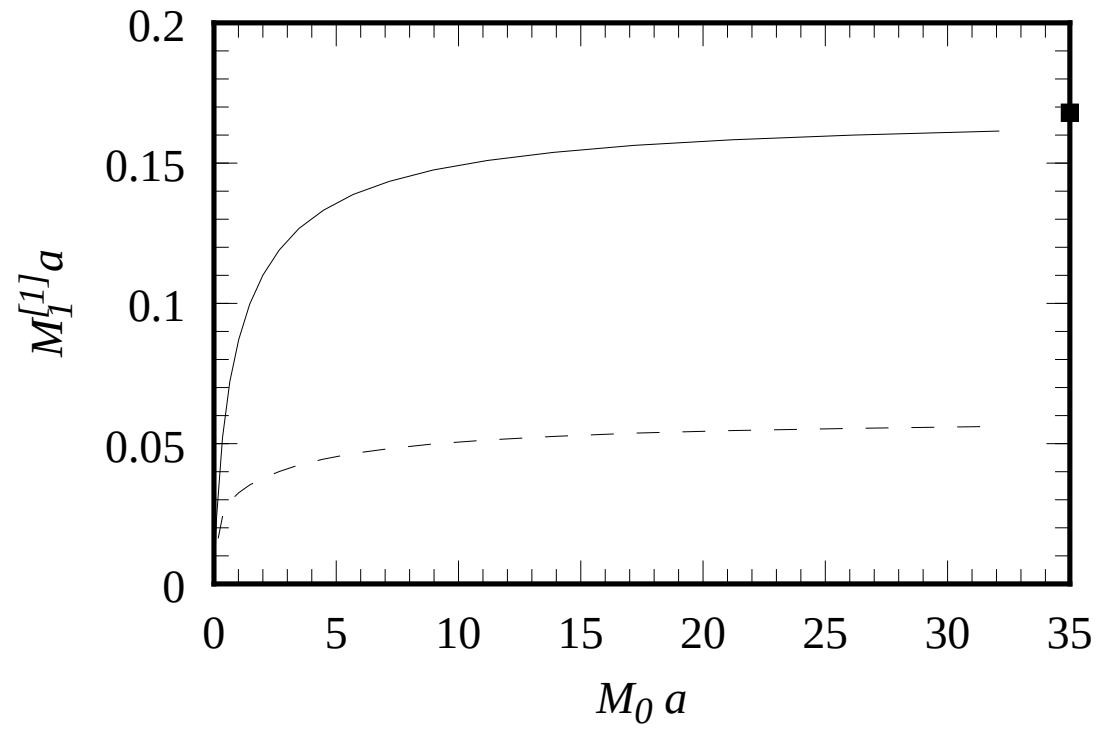

Figure 1. Plot comparing $M_1^{[1]}a$ vs. $M_0 a$ (solid curve) and $\tilde{M}_1^{[1]}a$ vs. $\tilde{M}_0 a$ (dashed curve). The static point is indicated by the box [5].

The third argument of $C$ denotes the parametric dependence. It is useful to take care of this term once and for all, and write

$$e^{M_1 a} = 1 + M_0 a + A_0 \sinh M_1 a - \widetilde{C} \tag{14}$$

where $M_0 a \equiv m_0 a - m_{0c} a = (2\kappa)^{-1} - (2\kappa_c)^{-1}$, and $\widetilde{C}(iM_1 a, \boldsymbol{0}; m_0 a) = C(iM_1 a, \boldsymbol{0}; m_0 a) - m_{0c} a$.

We turn now to one-loop results for $r = \zeta = 1$, $c_B = c_E = 0$, with and without tadpole improvement. In perturbation theory the rest mass has an expansion

$$M_1 a = \log(1 + \overset{(\sim)}{M}_0 a) + \sum_{l=1}^{\infty} g_0^{2l} \overset{(\sim)}{M}{}_1^{[l]} a. \tag{15}$$

In the tadpole improved version $\tilde{M}_0 a = M_0 a/u_0$, where $u_0$ is a suitable (gauge invariant) average link. In applications both $u_0$ and $\kappa_c$ would be taken from Monte Carlo calculations. Below we choose $u_0 = (8\kappa_c)^{-1}$. Figure 1 shows the one-loop correction to the rest mass $M_1$. As expected, $\tilde{M}_1^{[1]}$ is significantly smaller than $M_1^{[1]}$.

For the dynamic mass it is better to define a renormalization factor via $M_2 = m_2 Z_M$, where

$$m_2 a = \frac{M_0 a(1 + M_0 a)(2 + M_0 a)}{2\zeta^2(1 + M_0 a) + r\zeta M_0 a(2 + M_0 a)} \tag{16}$$

is the tree-level expression for the dynamic mass, except that the linear mass divergence is absorbed, order by order, into $M_0 a$. The tadpole improvement is $M_2 = \tilde{m}_2 \tilde{Z}_M$, where $\tilde{m}_2$ is given by the right-hand side of eq. (16), but with $M_0a \mapsto \tilde{M}_0a$. The factors $\overset{(\sim)}{Z}_M$ have series

$$\overset{(\sim)}{Z}_M = 1 + \sum_{l=1}^{\infty} g_0^{2l} \overset{(\sim)}{Z}{}_M^{[l]}, \tag{17}$$

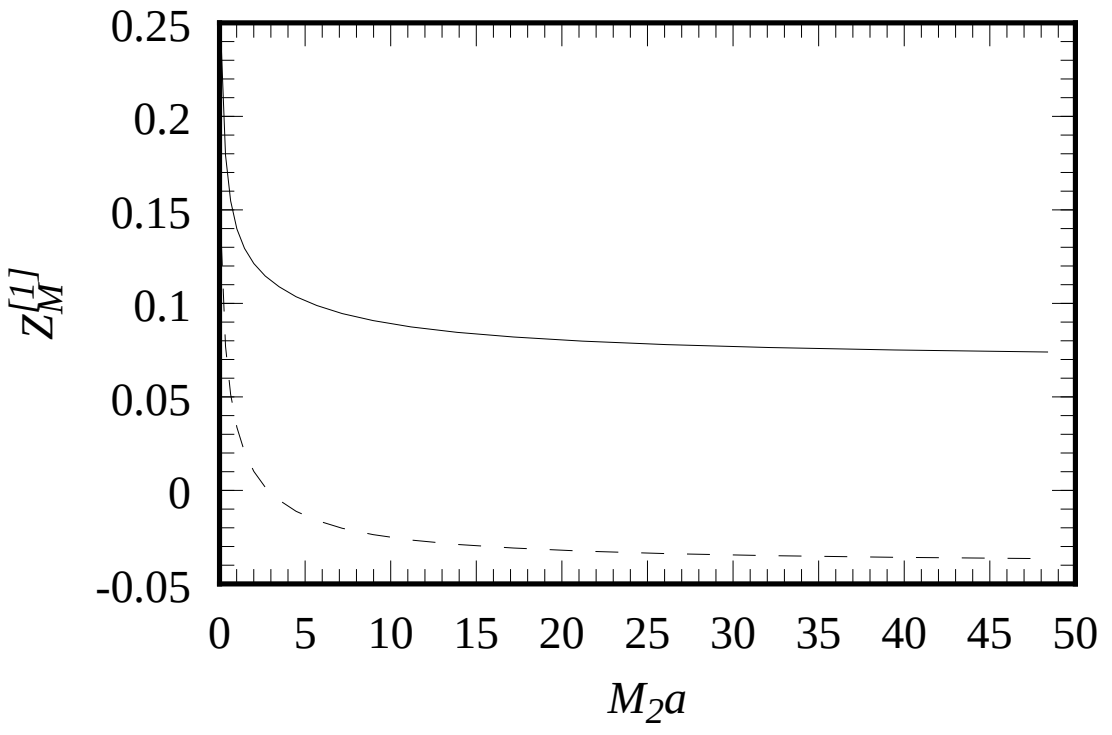

Figure 2. Plot comparing $Z_M^{[1]}$ (solid curve) and $\tilde{Z}_M^{[1]}$ vs. $M_2a$ (dashed curve).

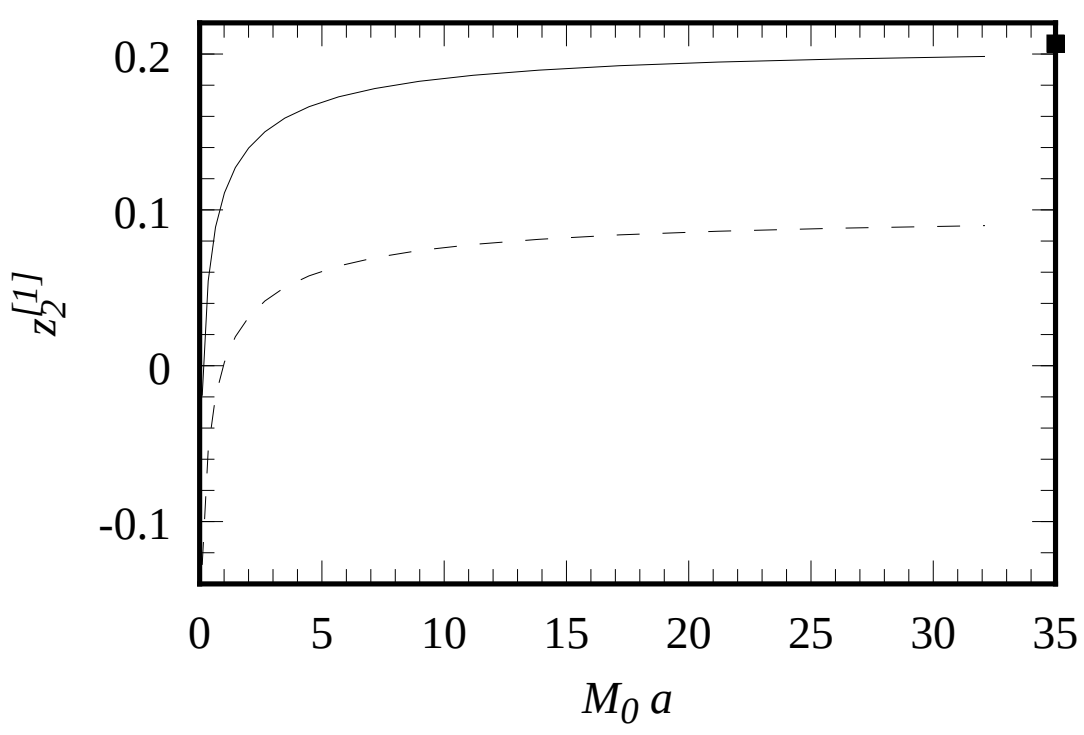

Figure 3. Plot comparing $z_2^{[1]}$ vs. $M_0a$ (solid curve) and $\tilde{z}_2^{[1]}$ vs. $\tilde{M}_0a$ (dashed curve). The static point is indicated by the box [5].

Figure 2 shows the one-loop renormalization of the dynamic mass. Again, $|\tilde{Z}_M^{[1]}|$ is significantly smaller than $|Z_M^{[1]}|$.

To define the perturbative coefficients for the wavefunction renormalization constant, factor out $e^{M_1a}$. The tadpole improved constant is $\tilde{Z}_2 = u_0 Z_2$. The perturbative series are

$$e^{M_1a}\overset{(\sim)}{Z}_2 = 1 + \sum_{l=1}^{\infty} g_0^{2l} \overset{(\sim)}{Z}{}_2^{[l]}. \tag{18}$$

The one-loop coefficient has an infrared divergence, which can be regulated with a gluon mass $\lambda$. Figure 3 plots the IR-finite $\overset{(\sim)}{z}{}_2^{[1]} = \overset{(\sim)}{Z}{}_2^{[1]} + \log(\lambda^2a^2)/(6\pi^2)$. Once again, $\tilde{z}_2^{[1]}$ is significantly smaller than $z_2^{[1]}$.

## 4. VERTEX CORRECTIONS

A full vertex function takes the form

$$V(p|q) = S(p)\Gamma(p|q)S(q). \tag{19}$$

In perturbation theory $\Gamma$ is given by the sum of all truncated three-point diagrams. To put the external lines on shell, one Fourier transforms in $p_0$ and $q_0$. Poles arise precisely as in the self-energy derivation, so the on-shell truncated vertex function is $\Gamma(iE_{\boldsymbol{p}}, \boldsymbol{p}|iE_{\boldsymbol{q}}, \boldsymbol{q})$. Consequently, when normalization conditions introduce a factor $1 + m_0a$ at tree level, the all-orders generalization is $e^{M_1a}$, just as with $Z_2^{-1}$.

## ACKNOWLEDGEMENTS

This work is being carried out in collaboration with Aida El-Khadra and Paul Mackenzie [1].

B.P.M. is supported in part by the U.S. Department of Energy under Grant No. DE-FG02-90ER40560. Fermilab is operated by Universities Research Association, Inc., under contract DE-AC02-76CH03000 with the U.S. Department of Energy.